\documentclass[a4paper,twocolumn,english,aps,prl,reprint,superscriptaddress,showpacs,showkeys,longbibliography]{revtex4-1}
\usepackage[latin9]{inputenc}
\usepackage{babel}
\usepackage{mathtools}
\usepackage{amsmath}
\usepackage{amssymb}
\usepackage{graphicx}
\usepackage[colorlinks=true, linkcolor=red, citecolor=blue, urlcolor=blue]{hyperref}
\usepackage[usenames,dvipsnames]{color}
\usepackage{bm}
\usepackage{bbold}
\usepackage{braket}
\usepackage{float}
\usepackage{comment}
\usepackage{dcolumn}

\definecolor{mygreen}{rgb}{0,0.5,0}
\definecolor{myblue}{rgb}{0,0,0.75}
\definecolor{mymagenta}{cmyk}{0,1,0,0.12}

\newcommand{\minus}{
  \setbox0=\hbox{-}
  \vcenter{
    \hrule width\wd0 height \the\fontdimen8\textfont3
  }%
}

\def\po{\tfrac{1}{2}}
\def\mo{\minus\tfrac{1}{2}}

\def\pl{\texttt{+}}
\def\mi{\texttt{-}}

\def\wt#1{\inner(#1)}
\def\inner(#1,#2,#3,#4,#5,#6){\ensuremath\left(\begin{array}{ccc} #1 & #2 & #3 \\ #4 & #5 & #6 \end{array}\right)}

\def\ws#1{\innerv(#1)}
\def\innerv(#1,#2,#3,#4,#5,#6){\ensuremath\left\{\begin{array}{ccc} #1 & #2 & #3 \\ #4 & #5 & #6 \end{array}\right\}}

\begin{document}

\title{Designing Frustrated Quantum Magnets with Laser-Dressed Rydberg Atoms}

\author{Alexander~W.~Glaetzle}
\email{alexander.glaetzle@uibk.ac.at}
\affiliation{Institute for Quantum Optics and Quantum Information of the Austrian
Academy of Sciences, A-6020 Innsbruck, Austria}
\affiliation{Institute for Theoretical Physics, University of Innsbruck, A-6020
Innsbruck, Austria}

\author{Marcello~Dalmonte}
\affiliation{Institute for Quantum Optics and Quantum Information of the Austrian
Academy of Sciences, A-6020 Innsbruck, Austria}
\affiliation{Institute for Theoretical Physics, University of Innsbruck, A-6020
Innsbruck, Austria}

\author{Rejish~Nath}
\affiliation{Institute for Quantum Optics and Quantum Information of the Austrian
Academy of Sciences, A-6020 Innsbruck, Austria}
\affiliation{Institute for Theoretical Physics, University of Innsbruck, A-6020
Innsbruck, Austria}
\affiliation{Indian Institute of Science Education and Research, Pune 411 008,
India}

\author{Christian~Gross}
\affiliation{Max-Planck-Institut f\"ur Quantenoptik, 85748 Garching, Germany}

\author{Immanuel~Bloch}
\affiliation{Max-Planck-Institut f\"ur Quantenoptik, 85748 Garching, Germany}
\affiliation{Fakult\"at f\"ur Physik, Ludwig-Maximilians-Universit\"at M\"unchen, 80799 Munich, Germany}

\author{Peter~Zoller}
\affiliation{Institute for Quantum Optics and Quantum Information of the Austrian
Academy of Sciences, A-6020 Innsbruck, Austria}
\affiliation{Institute for Theoretical Physics, University of Innsbruck, A-6020
Innsbruck, Austria}
\affiliation{Max-Planck-Institut f\"ur Quantenoptik, 85748 Garching, Germany}

\date{\today}
\begin{abstract}
We show how a broad class of lattice spin-1/2 models with angular- and distance-dependent couplings can be realized with cold alkali atoms stored in optical or magnetic trap arrays. The effective spin-1/2 is represented by a pair of atomic ground states, and spin-spin interactions are obtained by admixing van der Waals interactions between fine-structure split Rydberg states with laser light. The strengths of the diagonal spin interactions as well as the ``flip-flop", and ``flip-flip'' and  ``flop-flop'' interactions can be tuned by exploiting quantum interference, thus realizing different spin symmetries. The resulting energy scales of interactions compare well with typical temperatures and decoherence time scales, making the exploration of exotic forms of quantum magnetism, including emergent gauge theories and compass models, accessible within state-of-the-art experiments.
\end{abstract}



\maketitle

Understanding exotic forms of quantum magnetism is an outstanding
challenge of condensed matter physics~\cite{Lacroix:2011wn}.
Cold atoms stored in optical or magnetic trap arrays 
provide a unique platform to realize interacting quantum spins in
various lattice geometries with tunable interactions, and thus the
basic ingredients of competing magnetic orders and frustrated magnetism~\cite{lewenstein2012ultracold}.
A central experimental challenge for the observation of magnetic phases
with cold atoms is given by the requirement of ultralow temperatures
(and entropies), as set by the interaction scales of magnetic interactions.
For spin models derived from Hubbard dynamics for atoms in optical
lattices, this energy scale is set by the super-exchange processes,
$J\sim t_{H}^{2}/U$, with $t_{H}$ the hopping amplitude of atoms
between lattice sites, and $U$ the onsite interactions, resulting
in (rather small) energy scales of a few-tens of Hertz (or few nK)
regime~\cite{Trotzky:2008jya,*Greif:2013kb,*Anderlini:2007eb} (see, however, Ref.~\cite{Simon:2011hu,*Meinert:2013kl}). Instead, we consider below laser-excited interacting Rydberg
atoms~\cite{gallagher2005rydberg,*Saffman:2010wf,*Comparat:2010cba,*Low:2012cta}, which provide us not only with a complete toolbox to design and realize the complex spin-1/2
models of interest, but also give rise to energy
scales much larger than relevant decoherence rates. In contrast to models where a spin is encoded directly in a Rydberg state~\cite{Gunter:2013fv, *Lee:2013jk,*Muller:2008hb} we use ground state atoms weakly dressed with Rydberg states by laser light~\cite{Balewski:2014bc,*Hofmann:2013gma,*Pupillo:2010bta,*Henkel:2010ila,*Honer:2010jea,*Dauphin:2012joa,*Malossi:2014bl,*Glaetzle:2014vp}, which can be trapped in (large spacing) optical~\cite{Nelson:2007ks,*Schauss:2012hha,*Dudin:2012hma,*Nogrette:2014vk,*Schlosser:2011ef} or magnetic lattices~\cite{Abdelrahman:2010fh,*Leung:2014gw} of various geometries. {This provides a viable route to make phases of exotic quantum magnetism accessible to present atomic experiments.}

\begin{figure}[tb]
\centering \includegraphics[width=0.99\columnwidth]{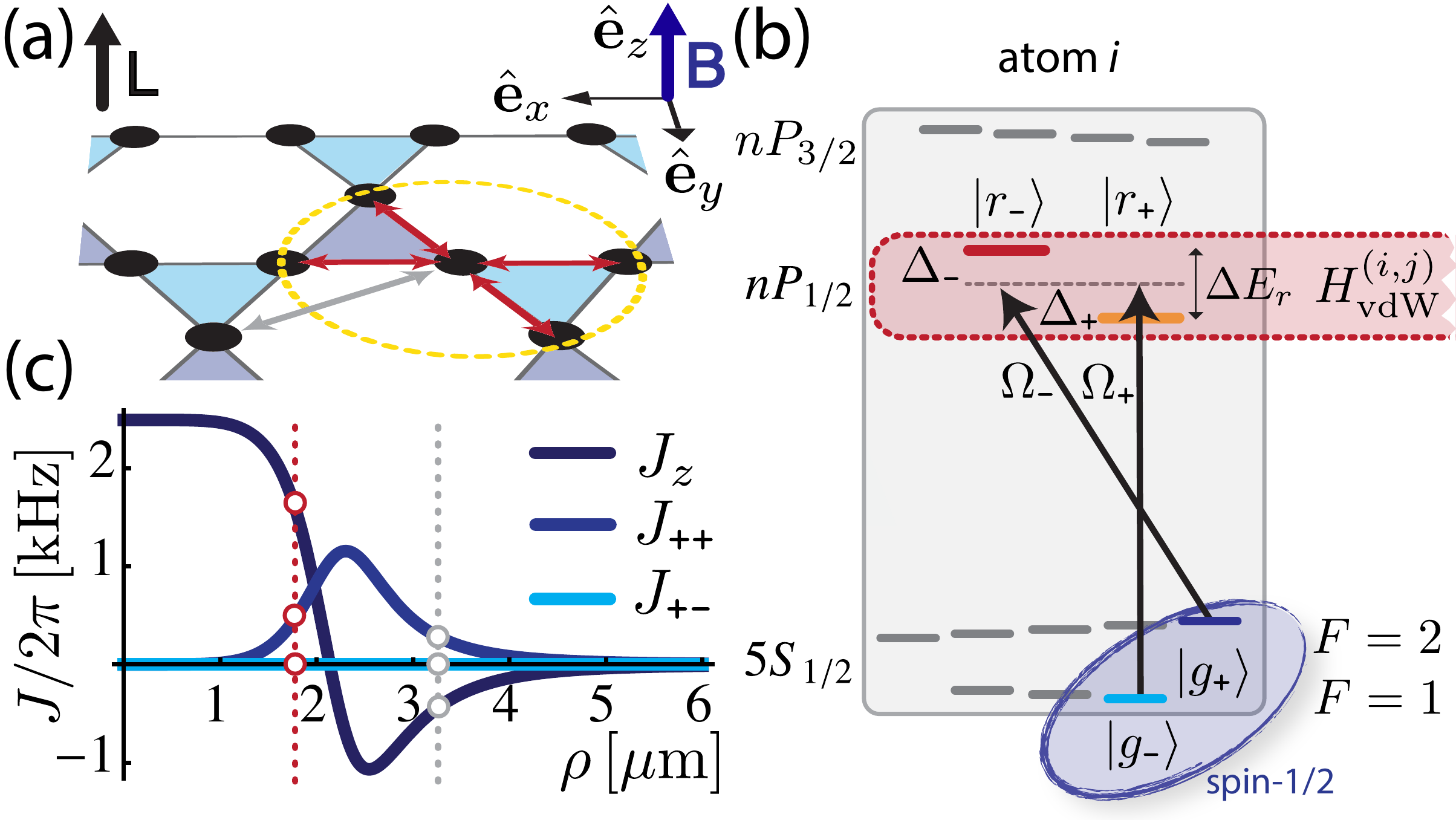}
\caption{{\small(a) Atoms loaded in a kagome lattice driven by laser light (L) propagating along the $z$-axis defined by the magnetic field (B).(b)~Atomic level scheme: ${}^{87}$Rb atoms with 
hyperfine ground states $|g_\sigma\rangle$ (representing  spin-1/2) coupled to $n^2P_{1/2}$ Rydberg states with $\sigma^{\pm}$ polarized  light and interacting via vdW interactions. (c)~Spin interactions $J_\alpha$ of Eq.~\eqref{eq:spinmodel} as a function of distance $\rho$ realizing quantum spin ice on a kagome lattice~\cite{Carrasquilla:2014vs}. Red (gray) dotted lines indicate NN and NNN interactions [red (gray) arrows in panel (a)]. Here $|r_{\pm}\rangle=|60^2P_{1/2},{\textstyle \pm\frac12}\rangle$ with Rabi frequencies \mbox{$\Omega_{\mi}=\Omega_\pl/4=2\pi\times 2.5$~MHz} and detunings \mbox{$\Delta_\mi=-\Delta_\pl=2\pi\times 50$~MHz} (so that  $J_{\pl\mi}=0$ --- see text).}}
\label{fig1} 
\end{figure}

We are interested in general $XYZ$ spin-1/2 models with both isotropic
and anisotropic interactions in 2D, as represented by the Hamiltonian
\begin{equation}
\begin{split}
H= & \sum_{\substack{i < j}
}\left[J_{z}(\mathbf{r}_{ij})S_{z}^{i}S_{z}^{j}+J_{||}(\mathbf{r}_{ij})S_{z}^{i}\right.\\
+ & \left.\frac{1}{2}\left(J_{\pl\mi}(\mathbf{r}_{ij})S_{\pl}^{i}S_{\mi}^{j}+J_{\pl\pl}(\mathbf{r}_{ij})S_{\pl}^{i}S_{\pl}^{j}+{\rm H.c.}\right)\right],\label{eq:spinmodel}
\end{split}
\end{equation}
where $S_{\alpha}^{j}$ are spin-1/2 operators at the lattice sites
$\mathbf{r}_{j}$. Our goal is to design spin-spin interaction patterns
$J_{\alpha}$, including nearest-neighbor (NN) and next-nearest-neighbor (NNN) couplings, as a function of $\mathbf{r}_{ij}=\mathbf{r}_{i}-\mathbf{r}_{j}$ including the range, angular dependence, and strength of the couplings.
Below we wish to illustrate the broad tunability offered by our setup
in the context of a paradigmatic example illustrated in Fig.~\ref{fig1}:
on a kagome lattice, different coupling realizations of Eq.~\eqref{eq:spinmodel}
encompass a variety of physical models, including kagome quantum spin ice
(requiring $J_{\pl\mi}=0$)~\cite{Carrasquilla:2014vs} and extended $XYZ$
models~\cite{He:2014ew,*Yan:2011kt}. These models encompass a prototypical feature of frustrated quantum magnets, i.e., the emergence of dynamical gauge fields~\cite{Lacroix:2011wn}. The specific form of the underlying gauge theories and the presence of topological spin liquid phases has been actively debated, making the controlled realization of such Hamiltonian dynamics timely matched with current theoretical efforts. {In the outlook we discuss further many-body perspectives.}

In our setup we consider single atoms loaded in trap arrays of
tunable geometry with spacings on the micrometer scale as demonstrated
in recent experiments~\cite{Nelson:2007ks, Schauss:2012hha,Dudin:2012hma, Nogrette:2014vk, Abdelrahman:2010fh,Leung:2014gw,Schlosser:2011ef}, {with negligible intersite tunneling}. We are interested here 
in alkali atoms, where a pair of states from the two hyperfine manifolds in the atomic ground
state represents the effective spin-1/2~\cite{Micheli:2006fq}. To be specific we consider
$^{87}$Rb atoms and choose %
\mbox{%
$|g_{\pl}\rangle\equiv|5^{2}S_{1/2},F=2,m_{F}=2\rangle$%
} and %
\mbox{%
$|g_{\mi}\rangle\equiv|5^{2}S_{1/2},F=1,m_{F}=1\rangle$%
} as our spin-1/2 [see Fig.~\ref{fig1}(b)].

Interactions between these effective spin states are induced by admixing
highly lying Rydberg states to the atomic ground states with laser
light, where van der Waals (vdW) interactions provide a strong coupling
even at micrometer distances. The key element is the excitation of
Rydberg states with finite orbital angular momentum exhibiting fine
structure splitting, and it is the combination of the spin-orbit interaction
and vdW interactions which provides the mechanism for the spin-spin
coupling. As indicated in Fig.~\ref{fig1}(b), we assume excitations
by left and right circularly polarized lasers with propagation direction
orthogonal to the lattice plane. In this configuration the ground states are coupled to
the two Rydberg Zeeman levels %
\mbox{$|r_{\sigma=\pm}\rangle\equiv|n^{2}P_{1/2},m_{j}=\pm\frac12\rangle\otimes|m_{I}={\textstyle \frac{3}{2}}\rangle$}. Here, 
\mbox{$|m_{I}={\textstyle \frac{3}{2}}\rangle$} is the maximally polarized nuclear spin state, which remains a spectator in our dynamics~\footnote{For laser detunings and vdW interactions much larger than the hyperfine splitting we can ignore the hyperfine interactions in the Rydberg state and the nuclear spin becomes a spectator.
}. This choice of laser configuration leads to spin couplings
$J_{\alpha}(\rho_{ij})$ with a \emph{purely radial} dependence as a function
of the distance $\rho_{ij}=|\mathbf{r}_{i}-\mathbf{r}_{j}|$, as shown
in Fig.~\ref{fig1}(c). This illustrates the design of kagome
quantum spin ice ($J_{\pl\mi}=0$) ~\cite{Carrasquilla:2014vs} for realistic atomic parameters. As discussed below, an angular dependence of $J_\alpha$ can be obtained
by inclining the laser beams~\cite{Reinhard:2007hma}.

\begin{figure}[tb]
\centering \includegraphics[width=0.99\columnwidth]{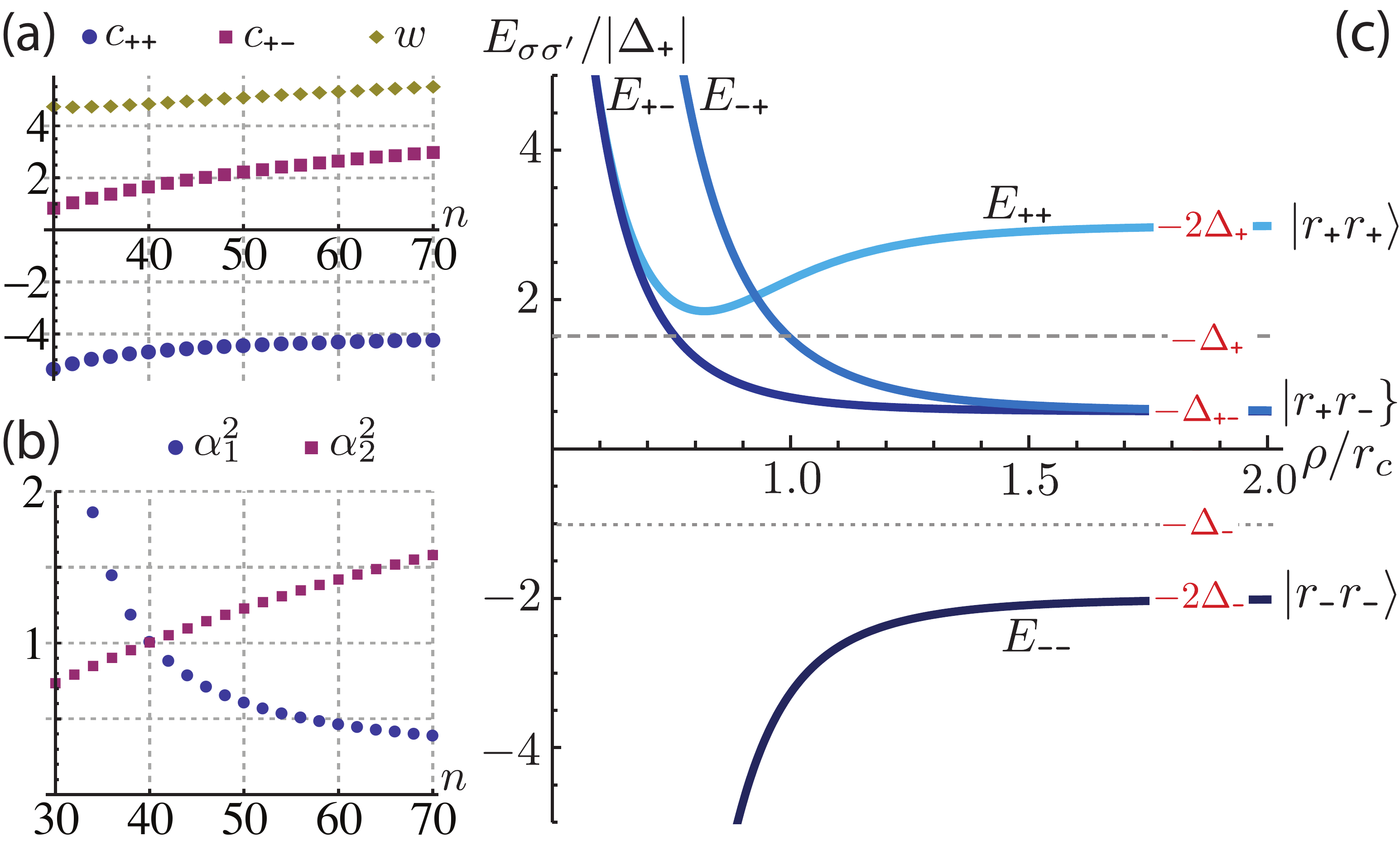}
\caption{{\small(a) $C_{6}$ coefficients of Eq.~\eqref{eq:HV-2} in atomic units for Rb atoms vs.~principal quantum number $n$.
(b) Ratios of diagonal and the $m$-changing
$C_6$. (c) Eigenenergies $E_{\sigma\sigma'}(\rho)$  (thick solid lines), energies of states
with a single Rydberg excitation in $r_{\pl}$ (gray dashed line)
and $r_{\mi}$ (gray dotted line) vs.~ $\rho/r_c$ with $r_c=(c_{\pl\pl}/2|\Delta_\pl|)^{1/6}$.
Asymptotic energies and eigenstates are indicated on the right. {Here, $\Delta_{\pl\mi}\equiv\Delta_{\pl}+\Delta_{\mi}$.}}}

{\small{}{}\label{fig2}} 
\end{figure}

To obtain the desired spin-spin interactions in Eq.~\eqref{eq:spinmodel} we consider a pair of atoms and
derive by adiabatic elimination of the Rydberg levels the
effective Hamiltonian for the ground state spins. Our starting point
is the microscopic Hamiltonian, $H_{{\rm mic}}=\sum_{i=1}^{2}\left[H_{A}^{(i)}+H_{L}^{(i)}\right]+H_{{\rm vdW}}$,
which is written as the sum of a single atom Hamiltonians including Zeeman split energy levels of the various states, the laser
driving and the vdW interaction. In the rotating frame we have $H_{A}^{(i)}=-\Delta_{\pl}|r_{\pl}\rangle_{i}\langle r_{\pl}|-\Delta_{\mi}|r_{\mi}\rangle_{i}\langle r_{\mi}|$
and $H_{L}^{(i)}  =\frac{1}{2}\Omega_{\pl}e^{i\varphi_{\pl}}|g_{\mi}\rangle_{i}\langle r_{\pl}|+\frac{1}{2}\Omega_{\mi}e^{i\varphi_{\mi}}|g_{\pl}\rangle_{i}\langle r_{\mi}|+{\rm H.c.}
$, where $\Delta_{\sigma}$ denotes the laser detunings, $\Omega_{\sigma}$
the Rabi frequencies and $\varphi_{\sigma}$ local laser phases. Since the derivation of the effective Hamiltonian is invariant under local gauge transformations, in the following we fix $\varphi_\sigma=0$ without loss of generality.

At the heart of our scheme is the vdW interaction $H_{{\rm vdW}}$
between the Zeeman sublevels in the $n^{2}P_{1/2}$ manifold. For the atomic configuration of Fig.~\ref{fig1}(a) (atoms in the $xy$-plane and lasers propagating along $z$)  this vdW interaction has the structure (see
SM) 
\begin{equation}
H_{{\rm vdW}}(\rho)=\left(\begin{array}{cccc}
V_{\pl\pl}(\rho) & 0 & 0 & W_{\pl\pl}(\rho)\\
0 & V_{\pl\mi}(\rho) & W_{\pl\mi}(\rho) & 0\\
0 & W_{\pl\mi}(\rho) & V_{\mi\pl}(\rho) & 0\\
W_{\pl\pl}(\rho) & 0 & 0 & V_{\mi\mi}(\rho)
\end{array}\right)\label{eq:HV-2}
\end{equation}
written in the basis $\{|r_{\mi}r_{\mi}\rangle,|r_{\mi}r_{\pl}\rangle,|r_{\pl}r_{\mi}\rangle,|r_{\pl}
r_{\pl}\rangle\}$ of Rydberg Zeeman states.
Here, $V_{\sigma\sigma'}({r})\equiv c_{\sigma\sigma'}/\rho^{6}$ are the (diagonal)
vdW interactions between the pair states $|r_\sigma r_{\sigma'}\rangle$,
with $V_{\pl\pl}=V_{\mi\mi}$ and $V_{\pl\mi}=V_{\mi\pl}$. In addition
we have ``flip-flop'' interactions between the states $|r_{\mi}r_{\pl}\rangle$
and $|r_{\pl}r_{\mi}\rangle$ and also ``flip-flip''
and ``flop-flop'' interactions between the Rydberg states $|r_{\mi}r_{\mi}\rangle$
and $|r_{\pl}r_{\pl}\rangle$, with coupling strength $W_{\pl\mi}$
and $W_{\pl\pl}$, respectively, where $W_{\pl\mi}(\rho)=-3W_{\pl\pl}(\rho)\equiv w/\rho^{6}$.
This arises from the fact that in our configuration the total magnetic
quantum number $M=m_{j}+m_{j'}$ can change by 0 or $\pm2$~\cite{Walker:2008tp}. The corresponding
$C_{6}$ coefficients $c_{\pl\pl}$, $c_{\pl\mi}$, and $w$ of Rb, which can be attractive or repulsive,  are plotted in Fig.~\ref{fig2}(a)
as a function of the principal quantum number $n$. We emphasize that in writing the time-independent Eq.~\eqref{eq:HV-2} we have assumed $\Delta E_r-(\Delta_\pl-\Delta_\mi)=0$, corresponding to the energy conservation condition for Raman processes between the spin ground states. We note, however, that $\Delta_\pl$ and $\Delta_\mi$ can still be chosen independently via an appropriate choice of laser frequencies and Rydberg Zeeman splitting $\Delta E_r$ \footnote{Away from the energy conservation condition the terms $W_{\pl\pl}$ in  \eqref{eq:HV-2} will carry a time-depencence \mbox{$\exp [i2(\Delta E_r-(\Delta_\pl-\Delta_\mi))t]$}}.

In the following we will derive the effective spin-spin
interactions of Eq.~\eqref{eq:spinmodel} by weakly admixing these Rydberg-Rydberg interactions
to the ground state manifold with lasers. We note, however, the basic
features of these spin-spin interactions can already be identified
in $H_{{\rm vdW}}$: (strong) diagonal interactions $V_{\pl\pl}$
and $V_{\mi\pl}$ will induce tunable $J_{z}$ interactions between
the dressed ground states $|g_{\mi}\rangle$ and $|g_{\pl}\rangle$,
while the couplings $W_{\pl\mi}$ and $W_{\pl\pl}$ give rise to
the $J_{\pl\mi}$ and $J_{\pl\pl}$ spin flip terms, respectively.

In the limit of weak laser excitation we obtain the  effective spin-spin interaction
between the ground states by treating the laser
interactions $H_{L}^{(1)}+H_{L}^{(2)}$ as a perturbation. As a first step we diagonalize
$H_{A}^{(1)}+H_{A}^{(2)}+H_{{\rm vdW}}$ as the dominant part of the
Hamiltonian in the subspace of two Rydberg excitations. {This results in four new eigenstates $ |E_{\sigma\sigma'}(\rho)\rangle$ and energies $E_{\sigma\sigma'}(\rho)$, ($\sigma, \sigma'=\pl,\mi$) shown in Fig.~\ref{fig2}(c), which can be
interpreted as Born-Oppenheimer adiabatic potentials (see the SM for a detailed discussion). We note that for
large distances $E_{\pl\pl}(\rho\rightarrow\infty)=-2\Delta_{\pl}$ and
$E_{\mi\mi}(\rho\rightarrow\infty)=-2\Delta_{\mi}$, corresponding to
states $|r_{\pl}r_{\pl}\rangle$ and $|r_{\mi}r_{\mi}\rangle$, respectively,
while the states $|r_{\mi}r_{\pl}\rangle$ and $|r_{\pl}r_{\mi}\rangle$
become asymptotically degenerate with energy $E_{\pl\mi}(\rho\rightarrow\infty)=E_{\mi\pl}(\rho\rightarrow\infty)=-(\Delta_{\pl}+\Delta_{\mi})$.}
The ratios $\alpha_{1}=W_{\pl\mi}/V_{\pl\mi}$ and $\alpha_{2}=W_{\pl\pl}/V_{\pl\pl}$
shown in Fig.~\ref{fig2}(b) determine the sign of the slope of the new eigenenergies at short distances. In particular, for $n^{2}P_{1/2}$
states of $^{87}$Rb we find that for $n\geq41$ the eigenenergies
$E_{\pl\pl}(\rho)$, $E_{\pl\mi}(\rho)$ and $E_{\mi\pl}(\rho)$ are repulsive
while $E_{\mi\mi}(\rho)$ is attractive at short distances [see Fig.~\ref{fig2}(c)]. For detunings $\Delta_\pl/\Delta_\mi<0$ and $\Delta_\pl+\Delta_\mi<0$ we avoid resonant Rydberg excitations for all distances; i.e.,~there are no zero crossings of $E_{\sigma\sigma'} (\rho)$, and perturbation theory in $\Omega_{\sigma}/|E_{\sigma'\sigma''}|$ is valid for all $\rho$. 

\begin{figure}[tb]
\centering \includegraphics[width=0.99\columnwidth]{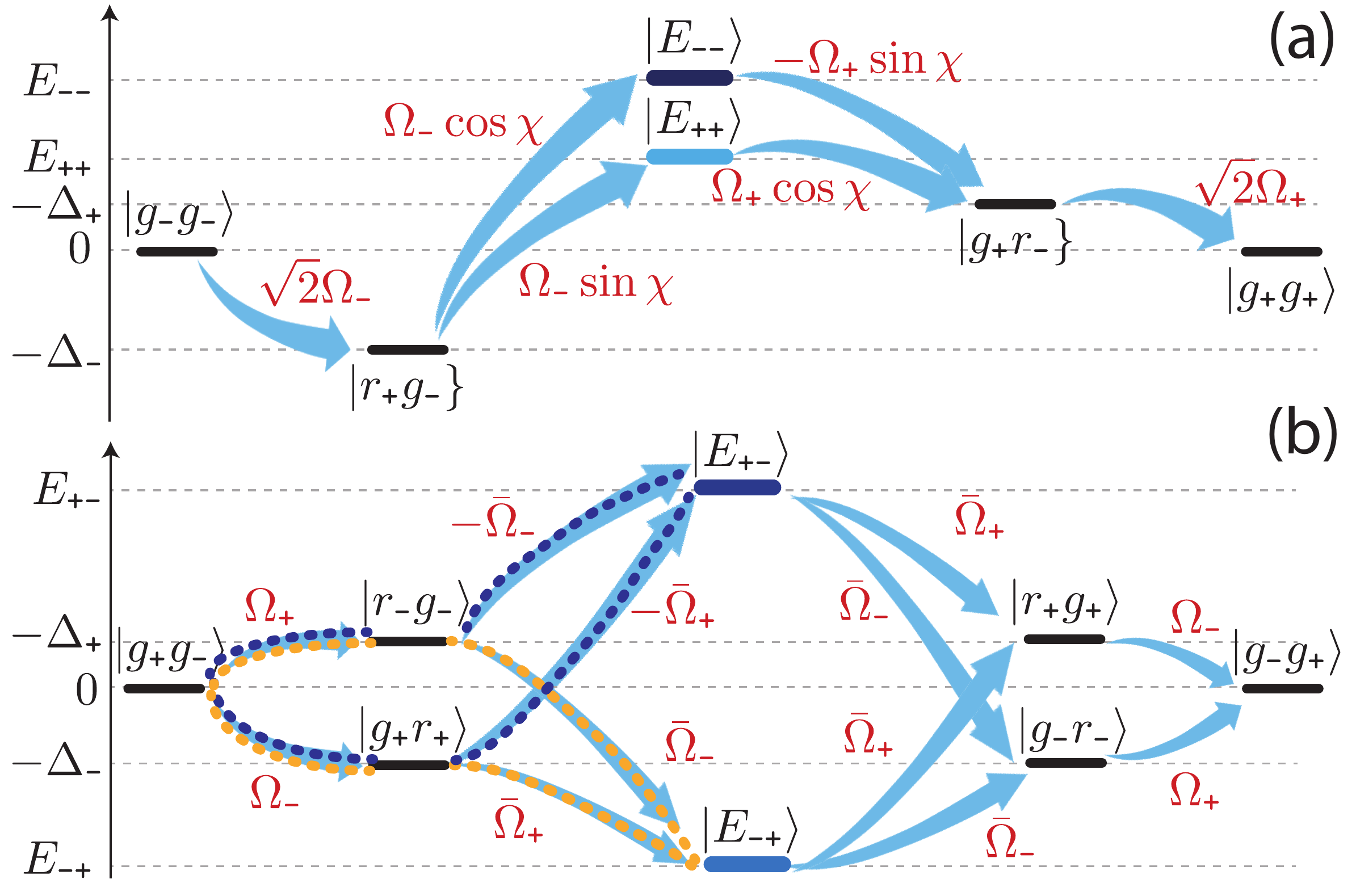}
\caption{{\small Path of perturbative couplings between the states (a) $|g_{\pl}g_{\pl}\rangle$
and $|g_{\mi}g_{\mi}\rangle$ and (b) $|g_{\pl}g_{\mi}\rangle$ and
$|g_{\mi}g_{\pl}\rangle$ visualizing the perturbative expressions behind $J_{\pl\pl}$
and $J_{\pl\mi}$ of Eqs.~\eqref{eq:J}, respectively. The energies $E_{\sigma\sigma'}$ are
plotted for a specific interatomic distance $\rho$ (with abbreviation $\bar\Omega_\sigma\equiv \Omega_\sigma/\sqrt{2}$). Yellow and blue dotted paths can interfere destructively (see text). }}
{\small{}{}{}\label{fig3} } 
\end{figure}

In fourth order, in the small parameter \mbox{$\Omega_{\sigma}/|E_{\sigma'\sigma''}|\ll1$} we obtain an effective spin-spin interaction Hamiltonian \mbox{$\tilde{H}=\sum_{\sigma,\sigma'}\left[\tilde{V}_{\sigma\sigma'}|g_{\sigma}g_{\sigma'}\rangle\langle g_{\sigma}g_{\sigma'}|+\tilde{W}_{\sigma\sigma'}|g_{\sigma}g_{\sigma'}\rangle\langle g_{\bar{\sigma}}g_{\bar{\sigma}'}|\right]$}
between the dressed ground states atoms. The diagonal interactions are
\begin{equation}
\begin{split}
\tilde{V}_{\sigma\sigma} & =\frac{\Omega_{\bar\sigma}^{4}}{8\Delta_{\bar\sigma}^{3}}\frac{V_{\pl\pl}\left(V_{\pl\pl}-2\Delta_{{\sigma}}\right)-W_{\pl\pl}^{2}}{W_{\pl\pl}^{2}-\left(V_{\pl\pl}-2\Delta_{\pl}\right)\left(V_{\pl\pl}-2\Delta_{\mi}\right)},\\
\tilde{V}_{\pl\mi} & =\frac{\Omega_{\pl}^{2}\Omega_{\mi}^{2}}{16\Delta_{\pl}^{2}\Delta_{\mi}^{2}}\left(\Delta_{\pl}+\Delta_{\mi}\right)\frac{V_{\pl\mi}\left(\Delta_{\pl\mi}-V_{\pl\mi}\right)+W_{\pl\mi}^{2}}{\left(\Delta_{\pl\mi}-V_{\pl\mi}\right)^{2}-W_{\pl\mi}^{2}},
\label{eq:tV}
\end{split}
\end{equation}
which, for small distances, are steplike potentials with $V_{\sigma\sigma'}(\rho\rightarrow0)=-\Omega_{\bar\sigma}^{2}\Omega_{\bar\sigma'}^{2}(\Delta_{\bar\sigma}+\Delta_{\bar\sigma'})/(16\Delta_{\bar\sigma}^2\Delta_{\bar\sigma'}^2)$. We have absorbed single particle light shifts in the definition
of the detunings (see Appendix). For the ``flip-flop'' and ``flop-flop'' interactions
we get 
\begin{equation}
\begin{split}\tilde{W}_{\pl\mi} & =\frac{\Omega_{\pl}^{2}\Omega_{\mi}^{2}}{16\Delta_{\pl}^{2}\Delta_{\mi}^{2}}\frac{\left(\Delta_{\pl}+\Delta_{\mi}\right)^{2}W_{\pl\mi}}{\left(\Delta_{\pl}+\Delta_{\mi}-V_{\pl\mi}\right)^{2}-W_{\pl\mi}^{2}},\\
\tilde{W}_{\pl\pl} & =-\frac{\Omega_{\pl}^{2}\Omega_{\mi}^{2}}{4\Delta_{\pl}\Delta_{\mi}}\frac{W_{\pl\pl}}{W_{\pl\pl}^{2}-\left(V_{\pl\pl}-2\Delta_{\pl}\right)\left(V_{\pl\pl}-2\Delta_{\mi}\right)},
\label{eq:tW}
\end{split}
\end{equation}
which are peaked at $R_{\pl\mi}^{6}=\sqrt{c_{\pl\mi}^2-w_{\pl\mi}^2}/|\Delta_{\pl}+\Delta_\mi|$
and $R_{\pl\pl}^{6}=\sqrt{(c_{\pl\pl}^2-w_{\pl\pl}^2)/(4\Delta_{\pl}\Delta_{\mi})}$,
respectively, and go to
zero for small and large distances. The spin couplings of Eq.~\eqref{eq:spinmodel} are then obtained as  
\begin{equation}
\begin{split} & J_{||}(\mathbf{r}_{ij})=\frac{1}{4}\left[\tilde{V}_{\pl\pl}(\mathbf{r}_{ij})-\tilde{V}_{\mi\mi}(\mathbf{r}_{ij})\right],\\
 & J_{z}(\mathbf{r}_{ij})=\frac{1}{4}\left[\tilde{V}_{\mi\mi}(\mathbf{r}_{ij})-2\tilde{V}_{\pl\mi}(\mathbf{r}_{ij})+\tilde{V}_{\pl\pl}(\mathbf{r}_{ij})\right],\\
 & J_{\pl\mi}(\mathbf{r}_{ij})=2\tilde{W}_{\pl\mi}(\mathbf{r}_{ij}),\,\text{and}\,{J}_{\pl\pl}(\mathbf{r}_{ij})=2\tilde{W}_{\pl\pl}(\mathbf{r}_{ij}).
 \label{eq:J}
\end{split}
\end{equation}
Figure~\ref{fig1}(c) shows a plot of Eq.~\eqref{eq:J} for $n=60$ $P_{1/2}${Rydberg states and for a particular set of laser parameters with \mbox{$\Delta_\mi=-\Delta_\pl$} such that  $J_{\pl\mi}=0$.} The diagonal $J_z$ interaction is steplike with a repulsive (antiferromagnetic) soft core at small distances, $\rho<2\;\mu$m, and an attractive (ferromagnetic tail) at long distances. The spin flip term $J_{\pl\pl}$ is peaked at $\rho\approx 2.5\,\mu$m while $J_{\pl\mi}=0$, thus realizing the Hamiltonian of quantum spin ice on a kagome lattice~\cite{Carrasquilla:2014vs} at a lattice spacing $a=1.8\;\mu$m. The lifetime of the $60P_{1/2}$ Rydberg state including blackbody radiation at $T=300$ K is $\tau_{60}=133\,\mu$s~\cite{Beterov:2009hna} which yields an effective ground state decay rate of $\Gamma_{{\rm eff}}=(\Omega_{\mi}/2\Delta_{\mi})^{2}\tau_{60}^{-1}\approx2\pi\times18$~Hz for $\Omega_{\mi}=2\pi\times5$~MHz and $\Delta_{\mi}=2\pi\times 50$~MHz, which is more than one to two orders of magnitude smaller than typical interaction energy scales shown in Fig.~\ref{fig1}(c). The fine structure splitting between the $60P_{1/2}$ and $60P_{3/2}$ manifolds is $\Delta E_{FS}\approx2\pi\times920$~MHz which is much larger than the Rydberg interactions for distances larger than about 2~$\mu$m.

The form and strength of the effective spin-spin interactions of Eqs.~\eqref{eq:tV} and~\eqref{eq:tW}  shown in Fig.~\ref{fig1}(c), including $J_{\pl\mi}=0$ for $\Delta_{\pl}=-\Delta_{\mi}$, can be understood in terms of {\em quantum interference} of the various paths contributing to the perturbation expressions of Eq.~\eqref{eq:J}.  These paths are illustrated in Fig.~\ref{fig3}:  both the states $|g_{\pl}g_{\pl}\rangle$ and $|g_{\mi}g_{\mi}\rangle$ [panel (a)] and also the states $|g_{\pl}g_{\mi}\rangle$ and $|g_{\mi}g_{\pl}\rangle$ [panel (b)] are coupled via four laser photons (blue arrows), giving rise to $\tilde{W}_{\pl\pl}$ and $\tilde{W}_{\pl\mi}$, respectively. In particular, the states $|g_{\pl}g_{\pl}\rangle$ and $|g_{\mi}g_{\mi}\rangle$ are coupled either via $|E_{\pl\pl}\rangle$ or via $|E_{\mi\mi}\rangle$ with position dependent coupling rates $\Omega_\sigma\sin\chi(\rho)$ and $\Omega_\sigma\cos\chi(\rho)$. For large distances $\sin\chi(\rho)\rightarrow0$ and thus $\tilde W_{\pl\pl}\rightarrow0$ while at short distances the ``flop-flop'' process is suppressed by the large resolvents $E_{\sigma\sigma'}^{-1}$ giving rise to the peaked form of $W_{\pl\pl}$ as a function of $\rho$. Panel (b) shows eight possible paths which can couple the $|g_{\pl}g_{\mi}\rangle$ and $|g_{\mi}g_{\pl}\rangle$ states. We note that both the two blue and the two yellow dotted paths coupling $|g_{\pl}g_{\mi}\rangle$ either to $|E_{\pl\mi}\rangle$ or to $|E_{\mi\pl}\rangle$, respectively, differ only by the energy denominators $\Delta_{\pl}^{-1}$ or $\Delta_{\mi}^{-1}$. Thus, for $\Delta_{\pl}=-\Delta_{\mi}$ the two yellow dotted paths and also the two blue dotted paths will {\em interfere destructively} and the ``flip-flop'' process vanishes, i.e.~$J_{\pl\mi}=2\tilde{W}_{\pl\mi}=0$, as shown in Fig.~\ref{fig1}(c).

\begin{figure}[tb]
\centering \includegraphics[width=0.99\columnwidth]{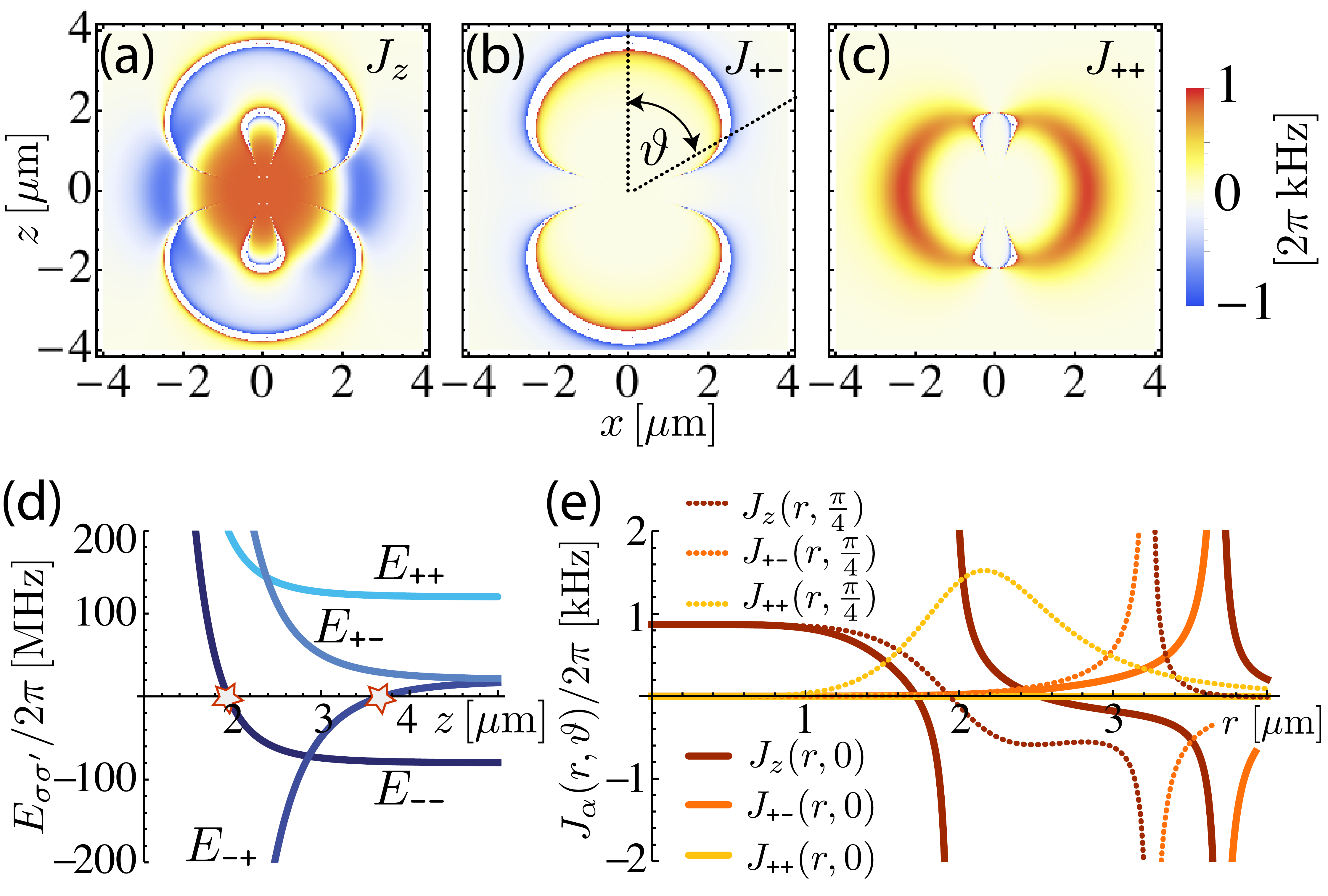}
\caption{\small 
(a)-(c) Effective spin-spin interactions $J_\alpha$ of Eq.~\eqref{eq:spinmodel} as a function of $r$ and $\vartheta$ for a laser propagating along the $z$ axes and atoms in the $zx$ plane. Here $|r_{\pm}\rangle=|60^2P_{1/2},{\textstyle \pm\frac12}\rangle$ with \mbox{$\Omega_{\mi}=\Omega_\pl/2=2\pi\times 5$~MHz} and \mbox{$(\Delta_\mi,\Delta_\pl)=2\pi\times (-60,40)$~MHz}. (d) Cut through the energy surfaces $E_{\sigma\sigma'}$ along the $z$ axis. In contrast to Fig.~\ref{fig2}(c) resonances appear, indicated with a star where $J_\alpha$ becomes singular as shown in panels (a)-(c). (e) Cut through panels (a)-(c) for $\vartheta=0$ (solid lines) and $\vartheta=\pi/8$ (dotted lines).}
\label{fig4} 
\end{figure}

We now turn to a setup with laser propagation direction (z axis) inclined with respect to the 2D plane containing the trapped atoms. This allows for an {\em angular dependence} (anisotropy) of the $J_{\alpha}(\mathbf{r}_{ij})$. In addition, we find as a new feature the appearance of {\em resonances} in the spin-spin couplings  as a function of spatial distance in the lattice. 
The origin of the anisotropy is the strong dependence of the various vdW interaction matrix elements on the angle $\vartheta$ between the  $z$ axis (defined by the laser propagation direction) and the relative vector connecting the two atoms $i$ and $j$ (see Appendix for details). As an example, we show in  Fig.~\ref{fig4} the spin-spin interactions for a propagation direction of both lasers parallel to the 2D plane ($zx$ plane). The anisotropy of the  $J_\alpha$ as a function of the angle $\vartheta$ is shown in panels (a)-(c). In particular, $W_{\pl\pl}(r,\vartheta)\sim\sin^{2}\vartheta$, and thus vanishes along the $z$ direction, reflecting the conservation of angular momentum  $M=m_j+m_j'$. In addition for $\vartheta\neq\pi/2$ resonances appear at specific interparticle distances, where one of the eigenenergies $E_{\sigma\sigma'}$ crosses the energy surface $E=0$ corresponding to ground state atoms $|g_{\sigma},g_{\sigma'}\rangle$ [indicated by the red stars in panel (d)]. This gives rise to clepsydra-shaped resonances in $J_{\alpha}$, as shown in panels (a)-(c), which in our perturbative treatment appear as singularities as a function of the distance, with $J_{\alpha}$ changing sign across the resonance.

We conclude with a perspective on the quantum many-body physics opened by the present work. The toolbox described above, together with techniques of adiabatic state preparation~\cite{Trebst:2006ga,*Kantian:2010kc,*Sorensen:2010gh} paves the way toward
the engineering of frustrated spin models, where different aspects of the interaction pattern can be exploited.
First, the independent tunability of both $J_{\pl\mi}$ and $J_{\pl\pl}$ couplings selects a 
particular spin symmetry, either conserving the total magnetization $\sum_j S^z_j$ or the parity $\prod_jS^z_j$, giving rise to a U(1) or $\mathbb{Z}_2$ symmetry, respectively. This finds immediate application 
in the context of extended quantum ice models~\cite{Huang:2014kq}, as  illustrated in Fig.~\ref{fig1} 
for kagome quantum spin ice~\cite{Carrasquilla:2014vs}. Within the same geometry, moving away from the 
$\Delta_\pl=-\Delta_\mi$ regime, a finite $J_{\pl\mi}$ can be switched on, and extended $XYZ$ models
can be realized~\cite{Gong:2014jz,He:2014ew,Yan:2011kt}. 
The ability of controlling each coupling strength in an angular- and 
distance-dependent way (c.f. Fig.~\ref{fig4}) points toward the realization of models displaying intermediate
symmetry, such as, e.g., compass models~\cite{Nussinov:2013wd}. By properly choosing the lattice spacings on a square lattice, it is possible
to single out interactions along one direction of pure $zz$ type, and of $\pl\pl$ type along the other, thus realizing extended square compass models. The large energy scales provided by the vdW interactions, combined with {\it in situ} measurement techniques demonstrated in large-spacing lattices~\cite{Nelson:2007ks,Schauss:2012hha,Dudin:2012hma,Nogrette:2014vk, Abdelrahman:2010fh,Leung:2014gw,Schlosser:2011ef}, have the potential to make the observation of different physical phenomena encompassed by these models, such as emergent gauge theories and exotic spin liquid states~\cite{Lacroix:2011wn}, accessible within Rydberg atom experiments.  Finally, we remark that the ideas proposed can be adapted to other dipolar systems such as, e.g.,~polar molecules~\cite{Wall2014}.

{\it Acknowledgments.} We thank A.~L\"auchli, M.~Lukin, R. Moessner, P.~Schau\ss, M.~Saffman, B.~Vermersch and  J.~Zeiher for stimulating discussions. This project was supported in part by the ERC Synergy Grant UQUAM, SIQS, RYSQ, COHERENCE, the SFB FoQuS (FWF Project No. F4016-N23), and the ERA-NET CHIST-ERA (R-ION consortium). 

{\it Note added.} --- In the final stages of the work we have been informed by T. Pohl of related work in the context of spin-1 models.~\cite{vanBijnen:2014wc}

\appendix
\section{Van der Waals interactions between $j=1/2$ Rydberg states}
Away from Foerster resonances two laser excited Rydberg atoms dominantly interact via van der Waals interactions~\cite{gallagher2005rydberg,*Saffman:2010wf,*Comparat:2010cba,*Low:2012cta}. These van der Waals interactions, $H_{\rm vdW}^{(i,k)}$, will mix different Zeeman sublevels $|m_j\rangle$ in the $nP_{1/2}$ manifold~\cite{Walker:2008tp}. Let us denote by \mbox{$\hat P=\sum_{i,j}| m_i,m_j\rangle\langle m_i,m_j|$} a projection operator into the $nP_{1/2}$ manifold, then dipole-dipole interactions 
\begin{equation*}
V_{\rm dd}^{(i,k)}(\mathbf{r})=-{\textstyle \sqrt{\frac{24 \pi}{5}} } \frac{1}{r^{3}} \sum_{\mu,\nu}C_{\mu,\nu;\mu+\nu}^{1,1;2}
 Y_{2}^{\mu+\nu}(\vartheta,\varphi)^* d^{(i)}_{\mu}d^{(j)}_{\nu},
\end{equation*}
will couple states in the $\hat P$ manifold to intermediate states, \mbox{$\hat Q_{\alpha,\beta}= |\alpha,\beta\rangle\langle \alpha,\beta|$}, which have an energy difference $\delta_{\alpha\beta}$. 
Here, $\mathbf{d}^{(i)}$ is the dipole operator of the  $i$-th atom and $\mathbf{r}=(r,\vartheta,\varphi)$ is the relative vector between atom $i$ and atom $j$ in spherical coordinates and $d^{(i)}_\mu$ is the $\mu$-th spherical components ($\mu,\nu\in \{-1,0,1\}$) of the atomic dipole operator. With $C_{m_1,m_2;M}^{j_1,j_2;J}$ we denote the Clebsch-Gordan coefficients and $Y_l^m$ are spherical harmonics. In second order perturbation theory this gives rise to
\begin{equation}
H_{\rm vdW}^{(i,k)}=\hat P\sum_{\alpha\beta}\frac{V_{\rm dd}^{(i,k)}\hat Q_{\alpha,\beta}V_{\rm dd}^{(i,k)}}{\delta_{\alpha\beta}}\hat P,
\label{eq:vdwop}
\end{equation}
where $H_{\rm vdW}^{(i,k)}$ is understood as an operator acting in the manifold of Zeeman sublevels.

\begin{figure*}[tb]
\centering
\includegraphics[width= 8 cm]{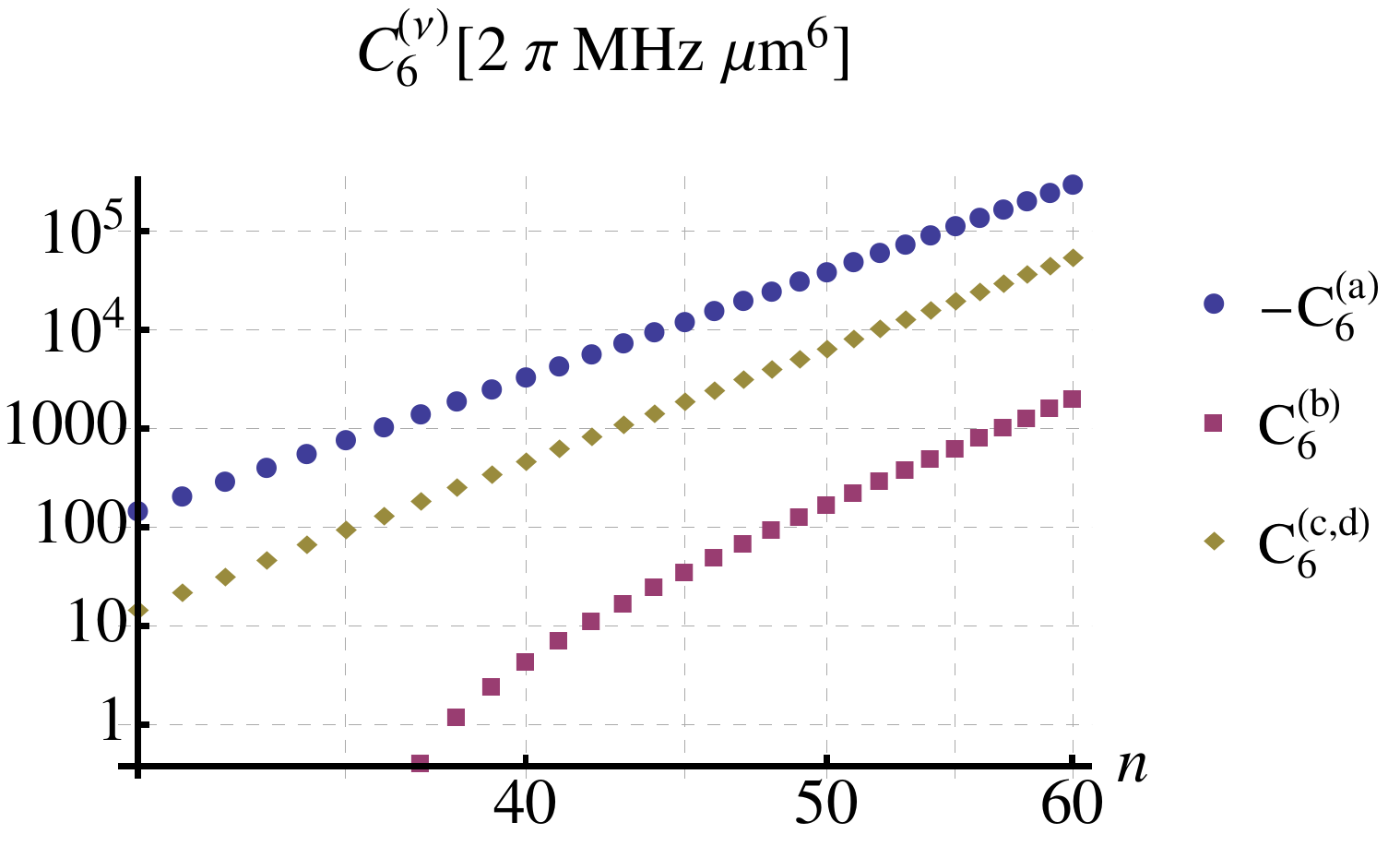}
\includegraphics[width= 8 cm]{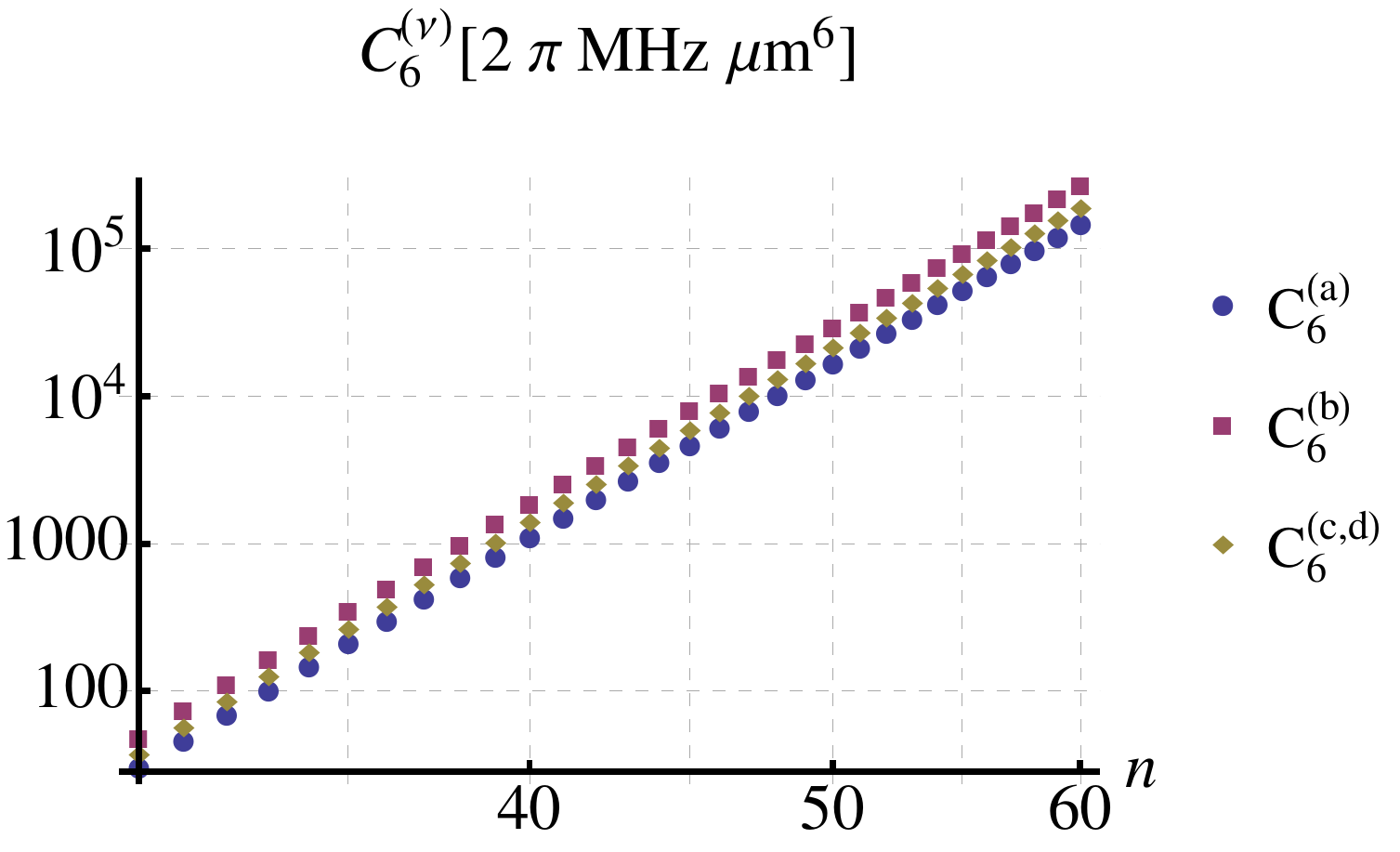}
\caption{\small{We plot the $C_6^{(\nu)}$ for (left) $nP_{1/2}$ and (right) $nS_{1/2}$ Rydberg states of $^{87}$Rb as a function of the principal quantum number $n$ for different channels $\nu$ of Tab.~\ref{tab:channels}.}}
\label{fig:c6ch} 
\end{figure*}

Due to the odd parity of the electric dipole operators $d^{(i)}_{\mu}$ and $d^{(j)}_{\nu}$, the dipole-dipole interaction, $V_{\rm dd}$, can couple initial $nP_{1/2}$ states only to $n'S_{1/2}$ or $n''D_{3/2}$ states. Therefore, there are four possible channels shown in~Tab.~\ref{tab:channels}(left) for which the matrix element $\langle nP_{1/2} m_1|\langle nP_{1/2} m_1|V_{\rm dd} |n',\ell_\alpha, j_\alpha,m_\alpha\rangle |n'',\ell_\beta, j_\beta,m_\beta\rangle$ of Eq.~\eqref{eq:vdwop} is non-zero. Here,  $(\ell_{\alpha,\beta}, j_{\alpha,\beta})$ can either correspond to $S_{1/2}$ or $D_{3/2}$ states depending on the channel.
While there is no selection rule for possible final principal quantum numbers $n'$ and $n''$ which solely determine the overall strength of the matrix element, the dipole-dipole matrix element is only non-zero if the magnetic quantum numbers and the spherical component of the dipole operator fulfill $m_1+\mu=m_\alpha$ and $m_2+\nu=m_\beta$.
The total vdW interaction of  Eq.~\eqref{eq:vdwop} can be obtained by summing over all channels $\nu$, that is 
\begin{equation}
\hat V_{\rm vdW}=\sum_{\nu} C_6^{(\nu)} \mathcal{D}_\nu(\vartheta,\varphi)/r^6. 
\end{equation}
Here, $C_6^{(\nu)}$ contains the radial part of the matrix elements 
\begin{equation}
C_6^{(\nu)}=\sum_{n_\alpha,n_\beta}\frac{\mathcal{R}_1^\alpha\mathcal{R}_2^\beta\mathcal{R}_3^\alpha\mathcal{R}_4^\beta}{\delta_{\alpha\beta}}
\end{equation}
which accounts for the overall strength of the interaction and is independent of the magnetic quantum numbers. With 
$\mathcal{R}_{i}^{k}=\int dr r^2 \psi_{n_i,\ell_i,j_i}(r)^* r\,\psi_{n_k,\ell_k,j_k}(r)$ we denote the radial integral.
The matrix
\begin{equation}
\mathcal{D}_\nu(\vartheta,\varphi)=\hat P\sum_{m_\alpha,m_\beta}\mathcal{M} \hat Q_{\alpha,\beta}^{(\nu)}\mathcal{M}\, \hat P
\end{equation}
on the other hand is a matrix in the subspace of magnetic quantum numbers which contains the relative angles between the two atoms ($s=1/2$)
\begin{widetext}
\begin{equation}
\begin{split}
\langle m_1,m_2|\mathcal{M}|m_\alpha,m_\beta\rangle=&
(-)^{s-m_1}\sqrt{(2\ell_1+1)(2j_1+1)(2\ell_\alpha+1)(2j_\alpha+1)}\ws{\ell_1,\ell_\alpha, 1,j_\alpha,j_1,s}\wt{\ell_\alpha,1,\ell_1,0,0,0}\\
\times&(-)^{s-m_2}\sqrt{(2\ell_2+1)(2j_2+1)(2\ell_\beta+1)(2j_\beta+1)}\ws{\ell_2,\ell_\beta, 1,j_\beta,j_2,s}\wt{\ell_\beta,1,\ell_2,0,0,0}\\
\times&\left[-\sqrt{\frac{24 \pi}{5}} \sum_{\mu,\kappa}C_{\mu,\kappa;\mu+\kappa}^{1,1;2} \wt{j_\alpha,1,j_1,m_\alpha,\mu,-m_1}\wt{j_\beta,1,j_2,m_\beta,\kappa,-m_2} 
 Y_{2}^{\mu+\kappa}(\vartheta,\varphi)^*\right],
\end{split}
\end{equation}
and $\hat Q_{\alpha,\beta}^{(\nu)}$ is a projector onto the final states $(\ell_{\alpha,\beta}, j_{\alpha,\beta})$ corresponding to channel $\nu\in\{a,b,c,d\}$ of Tab.~\ref{tab:channels}.
For the individual channels we find
\begin{equation}
\begin{split}
{\rm (a)}\quad&\mathcal{D}_a(\vartheta,\varphi)=\frac29\mathbb{1}_4-\mathcal{D}_0(\vartheta,\varphi),\\
{\rm (b)}\quad&\mathcal{D}_b(\vartheta,\varphi)=\frac49\mathbb{1}_4-\mathcal{D}_0(\vartheta,\varphi),\\
{\rm (c,d)}\quad&\mathcal{D}_c(\vartheta,\varphi)=\mathcal{D}_d(\vartheta,\varphi)=\mathcal{D}_0(\vartheta,\varphi),\\
\end{split}
\end{equation}
with $\mathbb{1}_4$ the $4\times 4$ identity matrix and 
\begin{equation}
\mathcal{D}_0(\vartheta,\varphi)=\frac{1}{81}\left(
\begin{array}{cccc}
 3 \cos (2 \vartheta )+11 & 3 e^{-i \phi }  \sin (2\vartheta ) & 3 e^{-i \phi }\sin (2\vartheta ) & 6 e^{-2 i \phi } \sin ^2(\vartheta ) \\
 3 e^{i \phi } \sin (2\vartheta ) & 13-3 \cos (2 \vartheta ) & -3 \cos (2 \vartheta )-5 & -3 e^{-i \phi } \sin (2\vartheta ) \\
 3 e^{i \phi } \sin (2\vartheta ) & -3 \cos (2 \vartheta )-5 & 13-3 \cos (2 \vartheta ) & -3 e^{-i \phi }  \sin (2\vartheta ) \\
 6 e^{2 i \phi } \sin ^2(\vartheta ) & -3 e^{i \phi } \sin (2\vartheta ) & -3 e^{i \phi } \sin (2\vartheta ) & 3 \cos (2 \vartheta )+11 \\
\end{array}
\right)\label{eq:D0}
\end{equation}
\end{widetext}
written in the basis $\{|\po \po\rangle,|\po\mo\rangle,|\mo\po\rangle,|\mo\mo\rangle\}$ of Zeeman states in the $j={1/2}$ Rydberg manifold. For the special orientations (i)~\mbox{$\vartheta=0$} we find
\begin{equation}
\mathcal{D}_0(0,0)=\frac{1}{81}\left(
\begin{array}{cccc}
 14 & 0 & 0 & 0 \\
 0 & 10 & -8 & 0 \\
 0 & -8 & 10 & 0 \\
 0 & 0 & 0 & 14 \\
\end{array}
\right)
\end{equation}
and (ii)~for $\vartheta=\pi/2$ the matrix simplifies to
\begin{equation}
\mathcal{D}_0({\textstyle\frac{\pi}{2}},0)=\frac{1}{81}\left(
\begin{array}{cccc}
 8 & 0 & 0 & 6 \\
 0 & 16 & -2 & 0 \\
 0 & -2 & 16 & 0 \\
 6 & 0 & 0 & 8 \\
\end{array}
\right).
\end{equation}
The total vdW interaction matrix in the $nP_{1/2}$ subspace becomes
\begin{multline}
H_{\rm vdW}^{(i,k)}=\left[\frac29 \left(C_6^{(a)}+2C_6^{(b)}\right)\mathbb{1}_4\right.\\
+\left.\left(2C_6^{(c)}-C_6^{(a)}-C_6^{(b)}\right)\mathcal{D}_0(\vartheta_{ik},\varphi_{ik})\right]/r_{ik}^6,
\label{eq:Hchannles}
\end{multline}
where the coefficients $C_6^{(\nu)}$ depend on the principal quantum number $n$, see Fig.~\ref{fig:c6ch}.

We note that the vdW Hamiltonian describing the interactions between $S_{1/2}$-states can be written in the exact same form as Eq.~\eqref{eq:Hchannles}. However, the coupling terms, $C_6^{(\nu)}$,  correspond to the channels of Tab.~\ref{tab:channels}(right). Therefore, the radial matrix elements for $S_{1/2}$ states differ only slightly due to the fine structure splitting $\Delta E_{FS}$ between $P_{1/2}$ and $P_{3/2}$ states, see Fig.~\ref{fig:c6ch}(right). In the limit where the fine structure can be neglected compared to other energy scales we find $C_6^{(a)}=C_6^{(b)}=C_6^{(c)}=C_6^{(d)}$ and the vdW interaction of Eq.~\eqref{eq:Hchannles} between $nS_{1/2}$ states becomes diagonal, that is $H_{\rm vdW}^{(i,k)}=(2/3) C_6^{(a)}\mathbb{1}_4$. Thus, there is no vdW mixing between Zeeman sublevels if the fine-structure splitting can be neglected. This can be understood by a simple argument: Since for $s$-states the different $m_j$ levels are proportional to the electronic spin $m_s$, that is $|m_j=\pm\frac12\rangle=|\ell=0,m_\ell=0\rangle\otimes |m_s=\pm\frac12\rangle$, and since dipole-dipole interactions cannot mix spin degrees of freedom there cannot be any vdW mixing of Zeeman levels in the absence of fine-structure. The first correction will be proportional to $\sim \Delta E_{FS}/\delta_{\alpha\beta}\mathcal{D}_0$.
It is therefore only the spin-orbit coupling in the intermediate $Q_{\alpha,\beta}$ manifold which mixes different Zeeman sublevels in the case of $S_{1/2}$ states. 

On the contrary, for $P_{1/2}$ states, the radial coefficients $C_6^{(\nu)}$ differ much more strongly due to the energy difference between $d$- and $s$-states and  due to the fact that Zeeman sublevels in the $nP_{1/2}$ manifold are already a superposition between $m_s=\pm\frac12$ states of the electronic spin. Therefore, mixing of Zeeman sublevels for $nP_{1/2}$ states can be of the same order of magnitude than the diagonal terms and play a significant role. In the special (1D) case $\vartheta=0$, the doubly excited levels $|\mo\mo\rangle$ and  $|\po\po\rangle$ are not coupled to any other doubly excited states which is a consequence of the conservation of the total angular momentum. On the contrary, for $\vartheta=\pi/2$ (atoms polarized perpendicular to the plane), the Hamiltonian of Eq.~\eqref{eq:Hchannles} reduces to Eq.~\eqref{eq:HV-2},
with
\begin{equation}
\begin{split}
&c_{\pl\pl}=\frac{2}{81}\left(5C_6^{(a)}+14 C_6^{(b)}+8 C_6^{(c)}\right),\\
&c_{\pl\mi}=\frac{2}{81}\left(C_6^{(a)}+10 C_6^{(b)}+16 C_6^{(c)}\right),\\
&w_{\pl\pl}=-\frac{2}{27}\left(C_6^{(a)}+ C_6^{(b)}-2 C_6^{(c)}\right),\\
&w_{\pl\mi}=\frac{2}{81}\left(C_6^{(a)}+ C_6^{(b)}-2 C_6^{(c)}\right)=-\frac{w_{\pl\pl}}{3}.
\end{split}
\end{equation}
shown in Fig.~\ref{fig2}(a) as a function of the principal quantum number $n$. In the following sections of this supplemental material, we will consider this particular orientation as it is the simplest configuration of vdW coupling where the doubly laser-excited state $|\mo\mo\rangle$ is only coupled to $|\po\po\rangle$.

\begin{table*}[tb]
\begin{center}
\begin{tabular}{llll}
\hline
$\nu$ &  $(\ell,j)+(\ell,j)$ & $\longrightarrow$ & $(\ell_\alpha,j_\alpha)+(\ell_\beta,j_\beta) $   \\
\hline
(a) &$P_{1/2}+ P_{1/2}$ & $\longrightarrow$ &$S_{1/2}+S_{1/2}$  \\	
(b) &$P_{1/2}+ P_{1/2}$ & $\longrightarrow$ &$D_{3/2}+D_{3/2}$  \\
(c) &$P_{1/2}+ P_{1/2}$ & $\longrightarrow$ &$S_{1/2}+D_{3/2}$  \\	
(d) &$P_{1/2}+ P_{1/2}$ & $\longrightarrow$ &$D_{3/2}+S_{1/2}$  \\
\hline
\end{tabular}
\qquad
\begin{tabular}{llll}
\hline
$\nu$ &  $(\ell,j)+(\ell,j)$ & $\longrightarrow$ & $(\ell_\alpha,j_\alpha)+(\ell_\beta,j_\beta) $   \\
\hline
(a) &$S_{1/2}+ S_{1/2}$ & $\longrightarrow$ &$P_{1/2}+P_{1/2}$  \\	
(b) &$S_{1/2}+ S_{1/2}$ & $\longrightarrow$ &$P_{3/2}+P_{3/2}$  \\
(c) &$S_{1/2}+ S_{1/2}$ & $\longrightarrow$ &$P_{1/2}+P_{3/2}$  \\	
(d) &$S_{1/2}+ S_{1/2}$ & $\longrightarrow$ &$P_{3/2}+P_{1/2}$  \\
\hline
\end{tabular}
\caption{Dipole-dipole interactions can couple $P_{1/2}$ (left) and $S_{1/2}$ (right) states to four channels (a-d). }
\label{tab:channels}
\end{center}
\end{table*}

\section{Laser excitation and hyperfine ground states}
\label{sec:single}

The laser Hamiltonian, $H_L^{(i)}$, couples two hyperfine ground states \mbox{$|g_\mi\rangle$} and \mbox{$|g_\pl\rangle$} to the Zeeman sublevels in the $nP_{1/2}$ Rydberg manifold with detunings $\Delta_\sigma$ and Rabi frequencies $\Omega_\sigma$ $(\sigma=\pl,\mi)$, respectively, see Fig.~\ref{fig1}(b). Uncoupling the nuclear spin the hyperfine ground states read
\begin{widetext}
\begin{equation*}
\begin{split}
&|g_\pl\rangle\equiv|5^2S_{1/2},F=2,m_F=2\rangle=|{\textstyle m_j= \frac{1}{2}}\rangle|m_I={\textstyle\frac{3}{2}}\rangle,\\
&|g_\mi\rangle\equiv|5^2S_{1/2},F=1,m_F=1\rangle=\frac{1}{2}\left[|{\textstyle m_j= \frac{1}{2}}\rangle|m_I={\textstyle\frac{1}{2}}\rangle-\sqrt{3}|m_j={\textstyle-\frac{1}{2}}\rangle|m_I={\textstyle\frac{3}{2}}\rangle\right],
\end{split}
\end{equation*}
where $m_I$ is the projection quantum number of the nuclear spin. Using $\sigma_\pl$ and $\sigma_\mi$ polarized light for the transition 
\begin{equation}
\begin{split}
&|g_\mi\rangle\xrightarrow{\Omega_\pl,\sigma_\pl} |nP_{1/2},m_j={\textstyle+\frac12}\rangle\otimes|m_I={\textstyle \frac32}\rangle,\\
&|g_\pl\rangle\xrightarrow{\Omega_\mi,\sigma_\mi} |nP_{1/2},m_j={\textstyle-\frac12}\rangle\otimes|m_I={\textstyle \frac32}\rangle, 
\end{split}
\end{equation}
respectively, couples to two different Rydberg states but both in the same nuclear state.  Thus, hyperfine structure can be treated as a spectator in the Rydberg manifold. Neglecting (small) hyperfine interactions, these are closed cycle transitions and do not couple to any other states in the hyperfine manifold. There are several alternative possibilities, e.g.
\begin{equation*}
\begin{split}
&|g_\pl\rangle\equiv|5^2S_{1/2},F=2,m_F=1\rangle= \frac{1}{2}\left[\sqrt{3}|{\textstyle m_j= \frac{1}{2}}\rangle|m_I={\textstyle\frac{1}{2}}\rangle+|m_j={\textstyle-\frac{1}{2}}\rangle|m_I={\textstyle\frac{3}{2}}\rangle\right],\\
&|g_\mi\rangle\equiv|5^2S_{1/2},F=1,m_F=0\rangle=\frac{1}{\sqrt{2}}\left[|{\textstyle m_j= \frac{1}{2}}\rangle|m_I={\textstyle-\frac{1}{2}}\rangle-|m_j={\textstyle-\frac{1}{2}}\rangle|m_I={\textstyle\frac{1}{2}}\rangle\right],
\end{split}
\end{equation*}
\end{widetext}
which can be laser excited to specific Rydberg states
\begin{equation*}
\begin{split}
|g_\mi\rangle\xrightarrow{\Omega_\pl,\sigma_\pl} |nP_{1/2},m_j={\textstyle-\frac12}\rangle\otimes|m_I={\textstyle \frac12}\rangle,\\
|g_\pl\rangle\xrightarrow{\Omega_\mi,\sigma_\mi} |nP_{1/2},m_j={\textstyle+\frac12}\rangle\otimes|m_I={\textstyle \frac12}\rangle.
\end{split}
\end{equation*}

\section{Effective ground state potentials}
\label{sec:adiabatic}

\subsection{Adiabatic elimination\label{sec:adi}}
In the dressing limit, $\Omega_\sigma\ll\Delta_{\sigma'}$,  atoms initially in their electronic ground states $|g\rangle_1\ldots|g\rangle_N$ are off-resonantelly coupled to the Rydberg states $|r\rangle_1\ldots|r\rangle_N$ and the new ``dressed'' ground states inherit a tunable fraction of the Rydberg interaction~\cite{Balewski:2014bc,*Hofmann:2013gma,*Pupillo:2010bta,*Henkel:2010ila,*Honer:2010jea,*Dauphin:2012joa,*Malossi:2014bl,*Glaetzle:2014vp}. The effective interaction potential between $N$ atoms in their dressed ground states, $|\tilde g\rangle_1\ldots|\tilde g\rangle_N$, can be obtained by diagonalizing the Hamiltonian $H_{\rm mic}$ for a fixed relative position and zero kinetic energy. The total Hamiltonian $H_{\rm mic}$  has block structure
\begin{equation}
\begin{split}
H_{\rm mic}&=\left(\begin{array}{ccccc}
\mathbf{H}_0 & \mathbf{\Omega}_1 & 0 & 0 &\\
\mathbf{\Omega}_1^\dag & \mathbf{H}_1 & \mathbf{\Omega}_2 & 0 &  \\
0 & \mathbf{\Omega}_2^\dag & \mathbf{H}_2 & \mathbf{\Omega}_3 & \\
0 & 0 & \mathbf{\Omega}_3^\dag & \mathbf{H}_3 &\\
 &  & &  & \ddots
 \end{array}\right)
 \end{split}
\end{equation}
where $\mathbf{H}_n$ governs the dynamics in the subspace with $n$-Rydberg excitations present, while the $\mathbf{\Omega}_n$ matrices describe the coupling between adjacent sectors $n$ and $n-1$ due to the laser. Only subspaces $\mathbf{H}_{n\geq 2}$ contain the interaction potentials $V_{ij}$ and $W_{ij}$ since we assume that ground and Rydberg states do not  interact at long distances.

{
As a first step we diagonalize
$H_{A}^{(1)}+H_{A}^{(2)}+H_{{\rm vdW}}$ as the dominant part of the
Hamiltonian in the subspace of two Rydberg excitations. This results
in four new eigenstates 
\begin{equation*}
\begin{split} & |E_{\sigma\sigma}(\rho)\rangle=\left[\cos\chi(\rho)|r_{\sigma}r_{\sigma}\rangle+\sigma\sin\chi(\rho)|r_{\bar{\sigma}}r_{\bar{\sigma}}\rangle\right]/\sqrt{2},\\
 & |E_{\sigma\bar{\sigma}}(\rho)\rangle=\left(|r_{\sigma}r_{\bar{\sigma}}\rangle+\sigma|r_{\bar{\sigma}}r_{\sigma}\rangle\right)/\sqrt{2}
\end{split}
\end{equation*}
where $\sigma=\pl,\mi$ and we defined $\bar\sigma\equiv-\sigma$ with corresponding eigenenergies 
\begin{equation}
\begin{split} & E_{\sigma\sigma}(\rho)=V_{\pl\pl}(\rho)-\Delta_{\pl\mi}+\sigma\sqrt{\delta_{\pl\mi}^{2}+W_{\pl\pl}(\rho)^{2}},\\
 & E_{\sigma\bar{\sigma}}(\rho)=V_{\pl\mi}(\rho)-\Delta_{\pl\mi}+\sigma W_{\pl\mi}(\rho),
 \label{eq:Esigma}
\end{split}
\end{equation}
shown in Fig.~\ref{fig2}(c). Here, we used the short hand notation $\Delta_{\pl\mi}\equiv\Delta_{\pl}+\Delta_{\mi}$, $\delta_{\pl\mi}\equiv\Delta_{\pl}-\Delta_{\mi}$ and  $\tan2\chi(\rho)=W_{\pl\pl}(\rho)/\delta_{\pl\mi}$. 
}

Adiabatically eliminating (up to fourth order in $\Omega_\sigma/E_{\sigma'\sigma''}\ll 1$) of the Rydberg states yields an effective interaction  in the subspace of hyperfine ground states 
\begin{equation}
\begin{split}
\tilde H&=\mathbf{H}_0+\mathbf{H}_1-\mathbf{\Omega}_1\mathbf{H}_1^{-1}\mathbf{\Omega}_1^\dag\\
&+\mathbf{\Omega}_1\mathbf{H}_1^{-1}\mathbf{\Omega}_1\mathbf{H}_1^{-1}\mathbf{\Omega}_1^\dag\mathbf{H}_1^{-1}\mathbf{\Omega}_1^\dag\\
&-\mathbf{\Omega}_1\mathbf{H}_1^{-1}\mathbf{\Omega}_2\mathbf{H}_2^{-1}\mathbf{\Omega}_2^\dag\mathbf{H}_1^{-1}\mathbf{\Omega}_1^\dag
\end{split}
\end{equation}
which yields (for two atoms)
\begin{equation}
\tilde H=\left(\begin{array}{cccc}\tilde  V_{\pl\pl} & 0 & 0 & \tilde W_{\pl\pl}  \\0 & \tilde V_{\pl\mi} & \tilde W_{\pl\mi} & 0 \\0 & \tilde W_{\pl\mi}^* & \tilde V_{\mi\pl} & 0 \\ \tilde W_{\pl\pl}^*  & 0 & 0 & \tilde V_{\mi\mi}\end{array}\right)
\end{equation}
written in the basis of the hyperfine states \mbox{$\{|g_{\pl}g_{\pl}\rangle,|g_{\mi}g_{\pl}\rangle,|g_{\pl}g_{\mi}\rangle,|g_{\mi}g_{\mi}\rangle\}$}. In the following we will discuss the various potentials separately.

\begin{widetext}
\subsection{The potential $\tilde V_{\pl\pl}$ and $\tilde V_{\mi\mi}$}
Adiabatic elimination up to fourth order in $\Omega/\Delta$ of the Rydberg states yields
\begin{equation*}
\begin{split}
&\tilde  V_{\pl\pl}=\frac{\Omega_\mi^2}{2 \Delta_\mi}-\frac{\Omega_\mi^4}{4 \Delta_\mi^3}+\frac{\Omega_\mi^4}{4 \Delta_\mi^2}\frac{ V_{\pl\pl}-2 \Delta_\pl}{ W_{\pl\pl}^2-\left(V_{\pl\pl}-2 \Delta_\mi\right) \left(V_{\pl\pl}-2 \Delta_\pl\right)},\\
&\tilde  V_{\mi\mi}=\frac{\Omega_\pl^2}{2 \Delta_\pl}-\frac{\Omega_\pl^4}{4 \Delta_\pl^3}+\frac{\Omega_\pl^4}{4 \Delta_\pl^2}\frac{ V_{\pl\pl}-2 \Delta_\mi}{ W_{\pl\pl}^2-\left(V_{\pl\pl}-2 \Delta_\mi\right) \left(V_{\pl\pl}-2 \Delta_\pl\right)},\\
\end{split}
\end{equation*}
where asymptotically we just recover the single particle light shifts (up to fourth order)
\begin{equation}
\begin{split}
&\tilde  V_{\pl\pl}^\infty\equiv\tilde  V_{\pl\pl}(r\rightarrow\infty)=\frac{\Omega_\mi^2}{2 \Delta_\mi}-\frac{\Omega_\mi^4}{8 \Delta_\mi^3},\\
&\tilde  V_{\mi\mi}^\infty\equiv\tilde  V_{\mi\mi}(r\rightarrow\infty)=\frac{\Omega_\pl^2}{2 \Delta_\pl}-\frac{\Omega_\pl^4}{8 \Delta_\pl^3}.
\end{split}
\end{equation}
The relative height of the potentials becomes
\begin{equation}
\begin{split}
&(\tilde  V_{\pl\pl}-\tilde  V_{\pl\pl}^\infty)/\tilde V_0=\frac{ 1-\alpha_{1}^2-\sigma\beta(r/R_1)^6}{ \alpha_{1}^2-\left[1-\sigma(r/R_1)^6\right] \left[1-\beta \sigma (r/R_1)^6\right]},\\
&(\tilde  V_{\mi\mi}-\tilde  V_{\mi\mi}^\infty)/\tilde V_0=\left(\frac{\Omega_\pl}{\Omega_\mi}\right)^4 \frac{1}{\beta^3} \frac{ 1-\alpha_{1}^2-\sigma(r/R_1)^6}{ \alpha_{1}^2-\left[1-\sigma(r/R_1)^6\right] \left[1-\beta \sigma (r/R_1)^6\right]},\\
\label{ref:relV11}
\end{split}
\end{equation}
with $\alpha_{1}^2=(w_{\pl\pl}/c_{\pl\pl})^2$ [shown in Fig.~\ref{fig2}(b)], $\beta=\Delta_\pl/\Delta_\mi$, $\tilde V_0=\Omega_\mi^4/(8 \Delta_\mi^3)$, $\sigma={\rm sign}(c_{\pl\pl}){\rm sign}(\Delta_\mi)$ and $R_{1}^6=|c_{\pl\pl}|/(2|\Delta_\mi|)$.
Due to the resolvent both potentials can be divergent for $W_{\pl\pl}^2-\left(V_{\pl\pl}-2 \Delta_\mi\right) \left(V_{\pl\pl}-2 \Delta_\pl\right)=0$, when two Born-Oppenheimer surfaces undergo an avoided crossing. This happens at
\begin{equation}
R_{\rm div}^6=\frac{\left(\Delta_\mi+\Delta_\pl\right) c_{\pl\pl}\pm\sqrt{\left(\Delta_\mi-\Delta_\pl\right)^2c_{\pl\pl}^2 +4 \Delta_\mi \Delta_\pl w_{\pl\pl}^2}}{4 \Delta_\mi \Delta_\pl}=\frac{1+\beta\pm\sqrt{\left(1-\beta\right)^2 +4 \beta\alpha_{1}^2}}{2 \beta }\sigma R_1^6.
\end{equation}
In order to avoid such divergences and to obtain steplike potentials we require ${\rm Im}(R_{\rm div})\neq 0$. For $\alpha_{1}^2>1$ this is for example the case when $\beta <1-2 \alpha_{1}^2 +2 \alpha_{1}\sqrt{\alpha_{1}^2 -1 }$.
\begin{figure*}[tb]
\centering
\includegraphics[width= 0.9\textwidth]{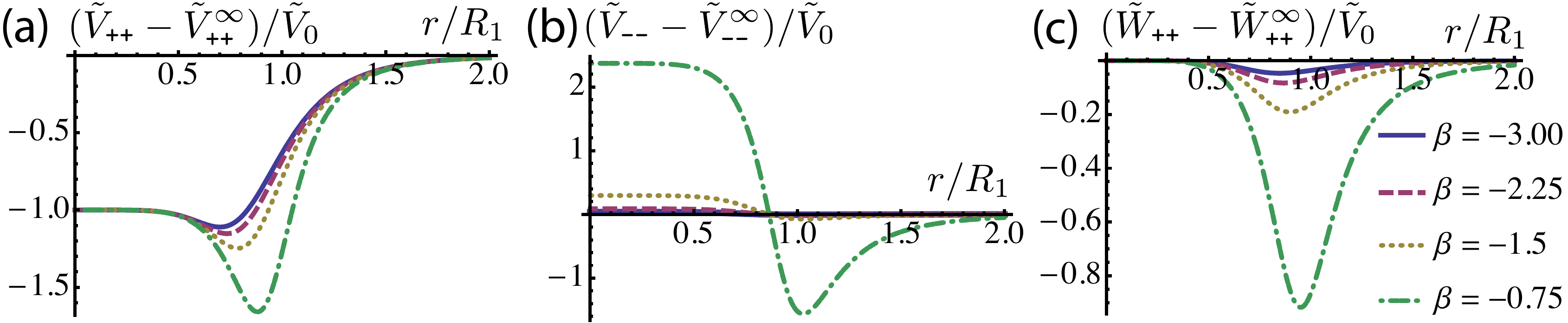}
\caption{\small{Plot of the relative height given by Eqs.~\eqref{ref:relV11} for $\alpha_{1}^2=1.41$ ($n=60$)  and $\sigma=-1$ for various laser detuning fractions $\beta$. }}
\label{fig:tildeV11} 
\end{figure*}
Figure~\ref{fig:tildeV11} shows a typical example of Eq.~\eqref{ref:relV11} for $n=60$ where $\alpha_{1}^2=1.41$ and $\sigma=-1$. In this case the potential has no singularity (avoided crossing) for $\beta<-0.30$.

For $\alpha_{1}=0$ one obtains the well known result of a single dressed Rydberg level, i.e. $(\tilde  V_{\pl\pl}-\tilde  V_{\pl\pl}^\infty)/\tilde V_0=-1/\left[1-\sigma(r/R_1)^6\right]$~\cite{Balewski:2014bc,*Hofmann:2013gma,*Pupillo:2010bta,*Henkel:2010ila,*Honer:2010jea,*Dauphin:2012joa,*Malossi:2014bl,*Glaetzle:2014vp}.

\subsection{Coupling element $\tilde W_{\pl\pl}$}
For the coupling matrix element $\tilde W_{\pl\pl}$ adiabatic elimination up to fourth order in $\Omega/\Delta$ yields
\begin{equation}
\begin{split}
&\tilde  W_{\pl\pl}=-e^{i \Delta \phi }\frac{ \Omega_\mi^2 \Omega_\pl^2 }{4 \Delta_\mi \Delta_\pl}\frac{W_{\pl\pl}}{ W_{\pl\pl}^2-\left(V_{\pl\pl}-2 \Delta_\mi\right) \left(V_{\pl\pl}-2 \Delta_\pl\right)}
\end{split}
\end{equation}
where $\Delta \phi = (\mathbf{k}_1 -  \mathbf{k}_2) (\mathbf{r}_1 + \mathbf{r}_2)$ is the phase difference between the two lasers at the center of mass position. Note that this phase can be gauged away using a local gauge transformation -- a rotation around the z-axis in the spin-basis. Asymptotically and at the origin ($r=0$) the coupling matrix element vanishes 
\begin{equation}
\begin{split}
\tilde  W_{\pl\pl}(r\rightarrow\infty)=0,\qquad\text{and}\qquad
\tilde  W_{\pl\pl}(r\rightarrow0)=0.
\end{split}
\end{equation}
In dimensionless units $\tilde W_{\pl\pl}$ reads
\begin{equation}
\begin{split}
\tilde  W_{\pl\pl}/\tilde V_0=
&-\frac{1 }{2\beta} \left(\frac{\Omega_\pl}{\Omega_\mi}\right)^2 \frac{\alpha_{1}\sigma (r/R_1)^6}{ \alpha_{1}^2-\left[1-\sigma (r/R_1)^6\right] \left[1-\beta \sigma (r/R_1)^6\right]}.
\label{eq:relW11}
\end{split}
\end{equation}
Again, this matrix element is regular for $\alpha_{1}^2>1$ and $\beta<1-2 \alpha_{1}^2 + 2  \alpha_{1}\sqrt{\alpha_{1}^2 -1 }$. 

Figure~\ref{fig:tildeV11}(c) shows a typical example of Eq.~\eqref{eq:relW11} for $n=60$ where $\alpha_{1}^2=1.41$ and $\sigma=-1$. In this case the potential has no singularity (avoided crossing) for $\beta<-0.30$. The coupling matrix element has a maximum at 
\begin{equation}
\begin{split}
&R_{1,{\rm max}}^6=\sqrt{(1-\alpha_{1}^2)/\beta}R_1^6\qquad \text{with}\\ &\tilde  W_{\pl\pl}(R_{1,{\rm max}})=-e^{i \Delta \phi}\frac{\tilde V_0 }{2\beta} \left(\frac{\Omega_\pl}{\Omega_\mi}\right)^2\frac{\alpha_{1} }{1+2 \sqrt{\beta\left(1-\alpha_{1}^2\right)}+ \beta }.
\end{split}
\end{equation}

\subsection{Potential $\tilde V_{\pl\mi}$}
For the potential $\tilde V_{\pl\mi}$ adiabatic elimination up to fourth order in $\Omega/\Delta$ yields
\begin{equation}
\begin{split}
&\tilde  V_{\pl\mi}=\frac{\Omega_\mi^2}{4 \Delta_\mi}+\frac{\Omega_\pl^2}{4 \Delta_\pl}-\frac{\Omega_\mi^4}{16 \Delta_\mi^3}-\frac{\Omega_\pl^2 \Omega_\mi^2}{16 \Delta_\mi^2 \Delta_\pl}-\frac{\Omega_\pl^2 \Omega_\mi^2}{16 \Delta_\mi \Delta_\pl^2}-\frac{\Omega_\pl^4}{16 \Delta_\pl^3}+\frac{\left(\Delta_\mi+\Delta_\pl\right)^2 \Omega_\mi^2 \Omega_\pl^2 \left(\Delta_\mi+\Delta_\pl-V_{\pl\mi}\right)}{16 \Delta_\mi^2 \Delta_\pl^2 \left(\left(\Delta_\mi+\Delta_\pl-V_{\pl\mi}\right)^2-W_{\pl\mi}^2\right)}\\
\end{split}
\end{equation}
where asymptotically we just recover the single particle light shifts (up to fourth order)
\begin{equation}
\tilde  V_{\pl\mi}^\infty\equiv\tilde  V_{\pl\pl}(r\rightarrow\infty)=\frac{\Omega_\mi^2}{4 \Delta_\mi}-\frac{\Omega_\mi^4}{16 \Delta_\mi^3}+\frac{\Omega_\pl^2}{4 \Delta_\pl}-\frac{\Omega_\pl^4}{16 \Delta_\pl^3}.
\end{equation}
The relative height of the potential becomes
\begin{equation}
(\tilde  V_{\pl\mi}-\tilde  V_{\pl\mi}^\infty)/\tilde V_0=\frac{\left(1+\beta\right) }{2\beta^2}\left(\frac{\Omega_\pl^2}{\Omega_\mi}\right)^2\frac{ \frac12(1+\beta) \sigma'(r/R_2)^6 -1+\alpha_{2}^2}{\left[\frac12 (1+\beta)\sigma'(r/R_2)^6-1\right]^2-\alpha_{2}^2}
\label{ref:relV12}
\end{equation}
\end{widetext}
\begin{figure*}[tb]
\centering
\includegraphics[width= 0.9\textwidth]{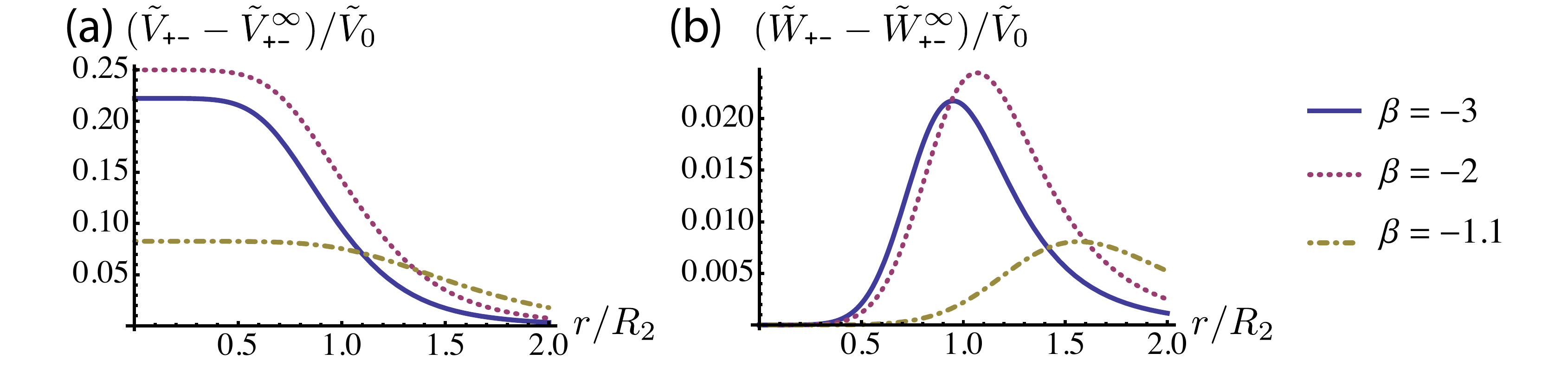}
\caption{\small{We plot the relative height of $\tilde V_{\pl\mi}$ given by Eq.~\eqref{ref:relV12} (left panel) and  of $\tilde W_{\pl\mi}$  given by Eq.~\eqref{eq:relW12} (right panel) for $\alpha_{2}^2=0.46$ ($n=60$), $\Omega_\mi=\Omega_\pl$  and $\sigma'=1$ for various laser detuning fractions $\beta$. }}
\label{fig:tildeV12} 
\end{figure*}
with $\alpha_{2}^2=(W_{\pl\mi}/c_{\pl\mi})^2$ [shown Fig.~\ref{fig2}(b)], $\sigma'={\rm sign}(c_{\pl\mi}){\rm sign}(\Delta_\mi)$ and $R_{2}^6=|c_{\pl\mi}|/(2|\Delta_\mi|)$.
Due to the resolvent the second term can be divergent for $\left(\Delta_\mi+\Delta_\pl-V_{\pl\mi}\right)^2-W_{\pl\mi}^2=0$, when two potential surfaces undergo an avoided crossing. This happens at
\begin{equation}
R_{\rm div}'^6=\frac{c_{\pl\mi}\pm w_{\pl\mi}}{\Delta_\mi+\Delta_\pl}=\frac{1\pm\alpha_{2}}{\frac12 (1+\beta)}\sigma' R_2.
\end{equation}
In order to avoid such divergences and to obtain steplike potentials we require $R_{\rm div}\in\mathbb{C}$. This can only be fulfilled for $-1<\alpha_{2}<1$ and $\beta<-1$. Figure~\ref{fig:tildeV12} shows a typical example of Eq.~\eqref{ref:relV12} for $n=60$ where $\alpha_{2}^2=0.46$ and $\sigma'=1$. In this case the potential has no singularity (avoided crossing) for $\beta<-1$. We note that for $\beta=-1$ the potential vanishes.

\subsection{Coupling element $\tilde W_{\pl\mi}$}
For the coupling matrix element $\tilde W_{\pl\mi}$ adiabatic elimination up to fourth order in $\Omega/\Delta$ yields
\begin{equation}
\begin{split}
&\tilde  W_{\pl\mi}=e^{i \Delta \phi_{12} }\frac{\Omega_\mi^2 \Omega_\pl^2}{16 \Delta_\mi^2 \Delta_\pl^2 }\frac{\left(\Delta_\mi+\Delta_\pl\right)^2 W_{\pl\mi}  }{\left(\Delta_\mi+\Delta_\pl-V_{\pl\mi}\right)^2-W_{\pl\mi}^2}
\end{split}
\end{equation}
where $\Delta \phi_{12} = (\mathbf{k}_1 -  \mathbf{k}_2) (\mathbf{r}_1 - \mathbf{r}_2)$ is the phase difference between the two lasers and relative position. Asymptotically and at the origin ($r=0$) the coupling matrix element vanishes 
\begin{equation}
\begin{split}
\tilde  W_{\pl\mi}(r\rightarrow\infty)=0,\qquad\text{and}\qquad
\tilde  W_{\pl\mi}(r\rightarrow0)=0.
\end{split}
\end{equation}
In dimensionless units $\tilde W_{\pl\mi}$ reads
\begin{equation}
\begin{split}
\tilde  W_{\pl\mi}/\tilde V_0=
\left(\frac{ \Omega_\pl}{  \Omega_\mi}\right)^2\frac{1}{2\beta^2}\frac{\alpha_{2}\frac12\left(1+\beta\right)^2 \sigma' (r/R_2)^6 }{\left[\frac12(1+\beta)\sigma'(r/R_2)^6-1\right]^2-\alpha_{2}^2}.
\label{eq:relW12}
\end{split}
\end{equation}
Again, this matrix element is regular for $-1<\alpha_{2}<1$ and $\beta<-1$.
Figure~\ref{fig:tildeV12} (right panel) shows a typical example of Eq.~\eqref{eq:relW12} for $n=60$ where $\alpha_{1}^2=1.41$ and $\sigma=-1$. We note that for $\beta=-1$ the coupling matrix element vanishes. The coupling matrix element has a maximum at 
\begin{equation}
\begin{split}
&R_{2,{\rm max}}^6=-\frac{2 \sqrt{1-\alpha_{2}^2 } }{1+\beta}R_2^6\qquad \text{with}\\ 
&\tilde  W_{\pl\mi}(R_{2,{\rm max}})=-\frac{\tilde V_0 }{4\beta^2} \left(\frac{\Omega_\pl}{\Omega_\mi}\right)^2\left(\sqrt{1-\alpha_{2}^2}-1\right) \frac{\beta +1}{ \alpha_{2} }.
\end{split}
\end{equation}

\bibliographystyle{apsrev4-1}
\bibliography{20141013.bib}

\end{document}